\documentclass[twocolumn,showpacs,preprintnumbers,amsmath,amssymb,widetext,super
scriptaddress]{revtex4}
\usepackage{graphicx}

\begin{document}

\title{Demonstration of an electrostatic-shielded cantilever.}

\author{P. Pingue}
\email{pingue@sns.it} 
\author{V. Piazza}
\affiliation{NEST-INFM \& Scuola Normale Superiore, I-56100 Pisa, Italy}
\author{P. Baschieri}
\author{C. Ascoli}
\affiliation{IPCF and CNR, I-56100 Pisa,Italy}
\author{C. Menozzi}
\author{A. Alessandrini} 
\author{P. Facci}
\affiliation{S3-INFM-CNR - Via Campi 213/A, I-41100 Modena, Italy}

\begin{abstract}
The fabrication and performances of cantilevered probes with reduced parasitic capacitance starting from a commercial $Si_{3}N_{4}$ cantilever chip is presented. Nanomachining and metal deposition induced by focused ion beam techniques were employed in order to modify the original insulating pyramidal tip and insert a conducting metallic tip. Two parallel metallic electrodes deposited on the original cantilever arms are employed for tip biasing and as ground plane in order to minimize the electrostatic force due to the capacitive interaction between cantilever and sample surface. Excitation spectra and force-to-distance characterization are shown with different electrode configurations. Applications of this scheme in electrostatic force microscopy, Kelvin probe microscopy and local anodic oxidation is discussed.
\end{abstract}
\maketitle

Scanning probe microscopy (SPM) is universally recognized as a powerful tool for the characterization of nanodevice surfaces. In particular electrostatic force microscopy (EFM) allows spatially resolved analysis of the sample-probe electrostatic interaction due for example to the electric-charge distribution\cite{Terris}, dopant profiles in semiconductors\cite{Abraham1}, local surface potentials\cite{Abraham2} (through the so-called Kelvin probe microscopy, KPM) and even the dielectric response of single molecules\cite{Gomes-Navarro}. Unfortunately a quantitative interpretation of EFM or KPM data is hindered by its complex dependence on the probe geometry. More in detail, electrostatic force interaction depends both on tip shape and on cantilever geometry. The cantilever contribution (which is normally few microns away from the sample surface) contributes as a large additional capacitance and gives rise to a background force on top of which the tip-sample interaction has to be detected. For this reason its effect on the total force gradient must be taken into account for a thorough understanding of the experimental data.

\begin{figure}[ht!]
\begin{center}
\includegraphics[width=8 cm,clip]{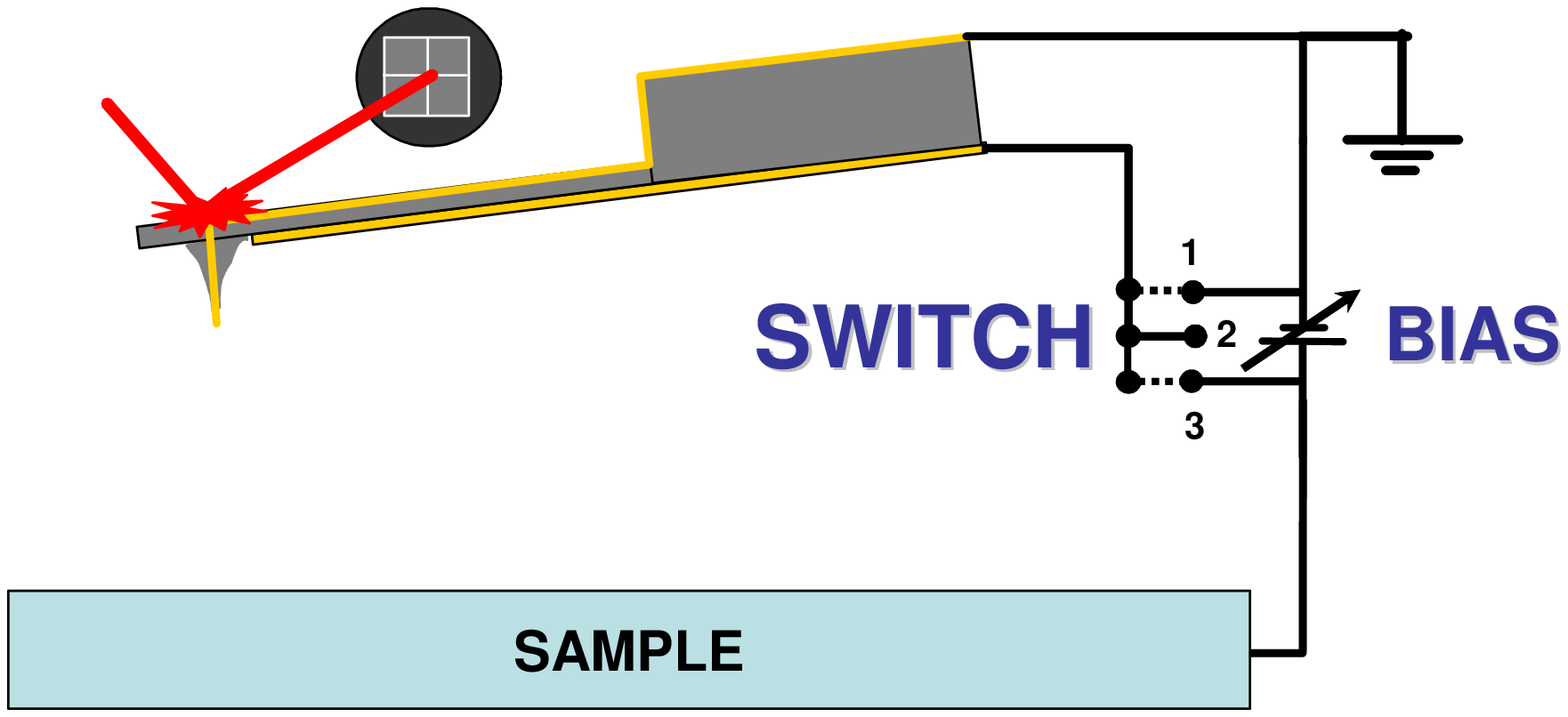}
\includegraphics[width=8 cm,clip]{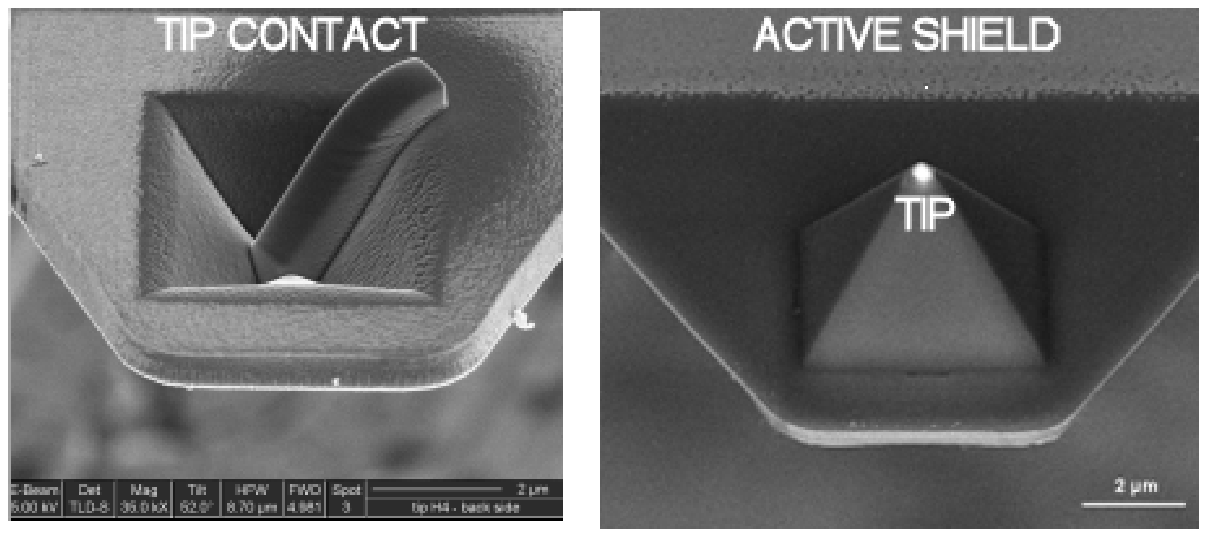}
\end{center}
\caption{(a) Scheme of the probe;(b) SEM images of the FIB-modified cantilever apex.}
\label{fig1}
\end{figure}

In standard EFM a bias voltage is applied between a conductive tip (and cantilever) and the sample under study. This bias induces electric fields which are related to the charge distribution on the sample and strongly depend on its conductivity. A number of solutions to such electrostatic configurations were proposed by several groups\cite{Baro,Saenz,Law,Koley} in order to figure out the role of the cantilever in the resolution of EFM imaging. The general results of these theoretical studies show that the tip apex contribution is significant only in the sub 10 nm range of tip-sample distance, whereas the tip cone and the cantilever capacitances dominate above 10 nm and 100 nm respectively. In particular, for a triangular cantilever and tip length of few microns, the capacitance gradient ratio between the cantilever and the tip is larger than one starting from few tens of nanometers\cite{Baro,Koley}. This ratio controls the accuracy of the EFM measurements. For example, measurements in integrated circuits are often done at large tip-to-sample distance when operating on a probe station or when a circuit passivation layer is present. In this situations tip-cone and cantilever interactions with the sample under analysis are not negligible and lead to a drastic reduction of the spatial resolution achievable. Probes having the smallest cantilever area are desirable to improve the EFM and KPM resolution and accuracy\cite{Baro,Koley}.
Moreover, a probe with reduced tip-cone to sample capacitance has been previously fabricated using an innovative procedure based on Focused Ion Beam (FIB) nanolithographic technique starting from a standard gold-coated $Si_{3}N_{4}$ triangular cantilever\cite{Menozzi}. The main feature of this probe consists in a FIB-deposited Pt electrode, that allows to reduce the tip-cone to sample capacitance. A reduced gold coating on the top side of the cantilever was present as electrical connection to the Pt stripe for tip biasing. Fabrication details can be found elsewhere\cite{Menozzi}. Conductivity and reduced cone to sample capacitance of this probe were demonstrated in Ref.\cite{Menozzi}, where current sensing AFM on gold film was reported. At a fixed distance (15 nm) from the sample surface, the electrostatic interaction at different voltages of the FIB-modified probe was compared with a standard gold-coated $Si_{3}N_{4}$ cantilever, obtaining a significant decrease of the electrostatic force interaction in that range of distances. Moreover, a reduction of the width of the gold-coated cantilever stripe allowed in that case a partial reduction of the electrostatic interaction of the cantilever itself with the substrate.

\begin{figure}[ht!]
\begin{center}
\includegraphics[width=8 cm,clip]{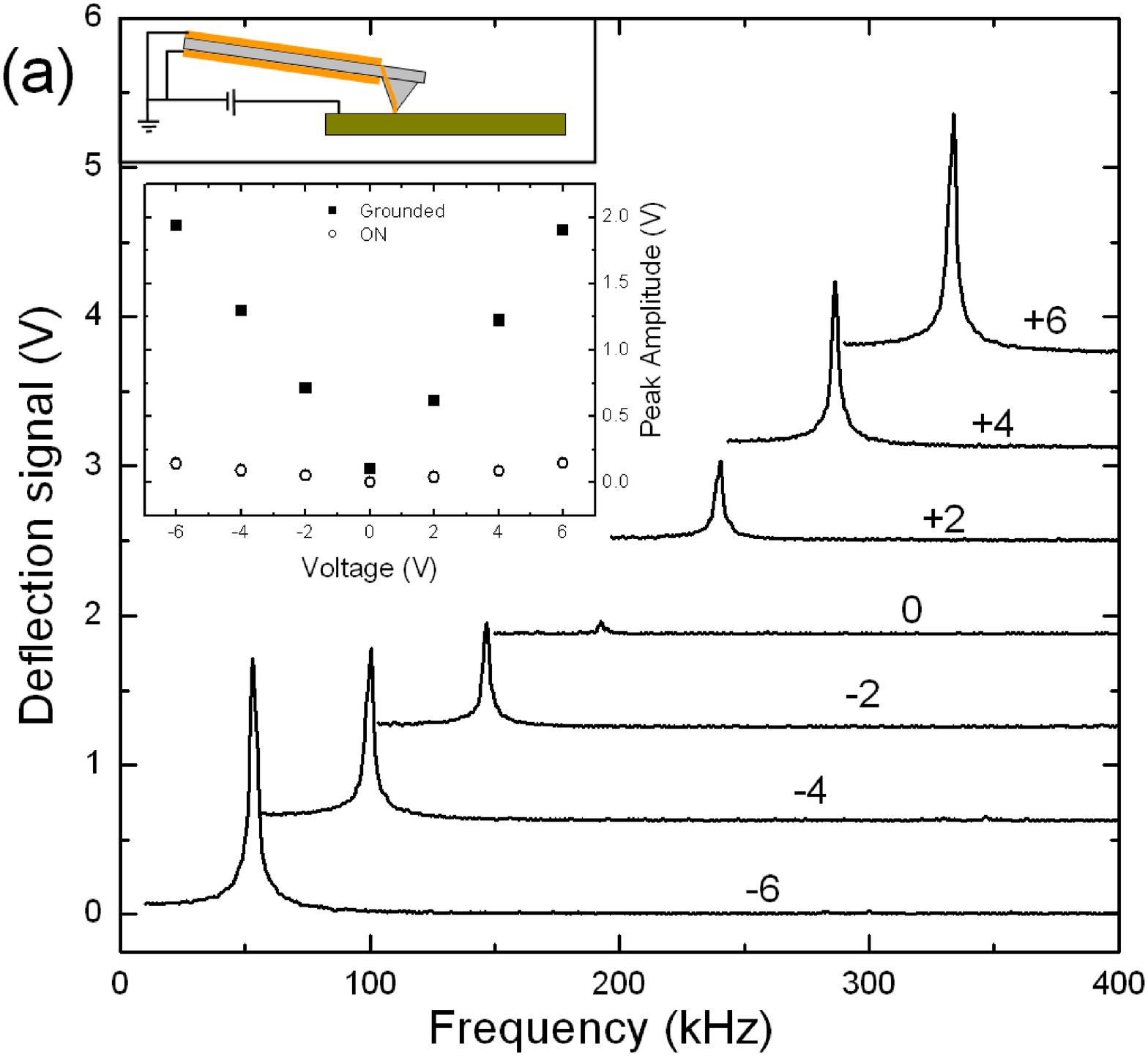}
\includegraphics[width=8 cm,clip]{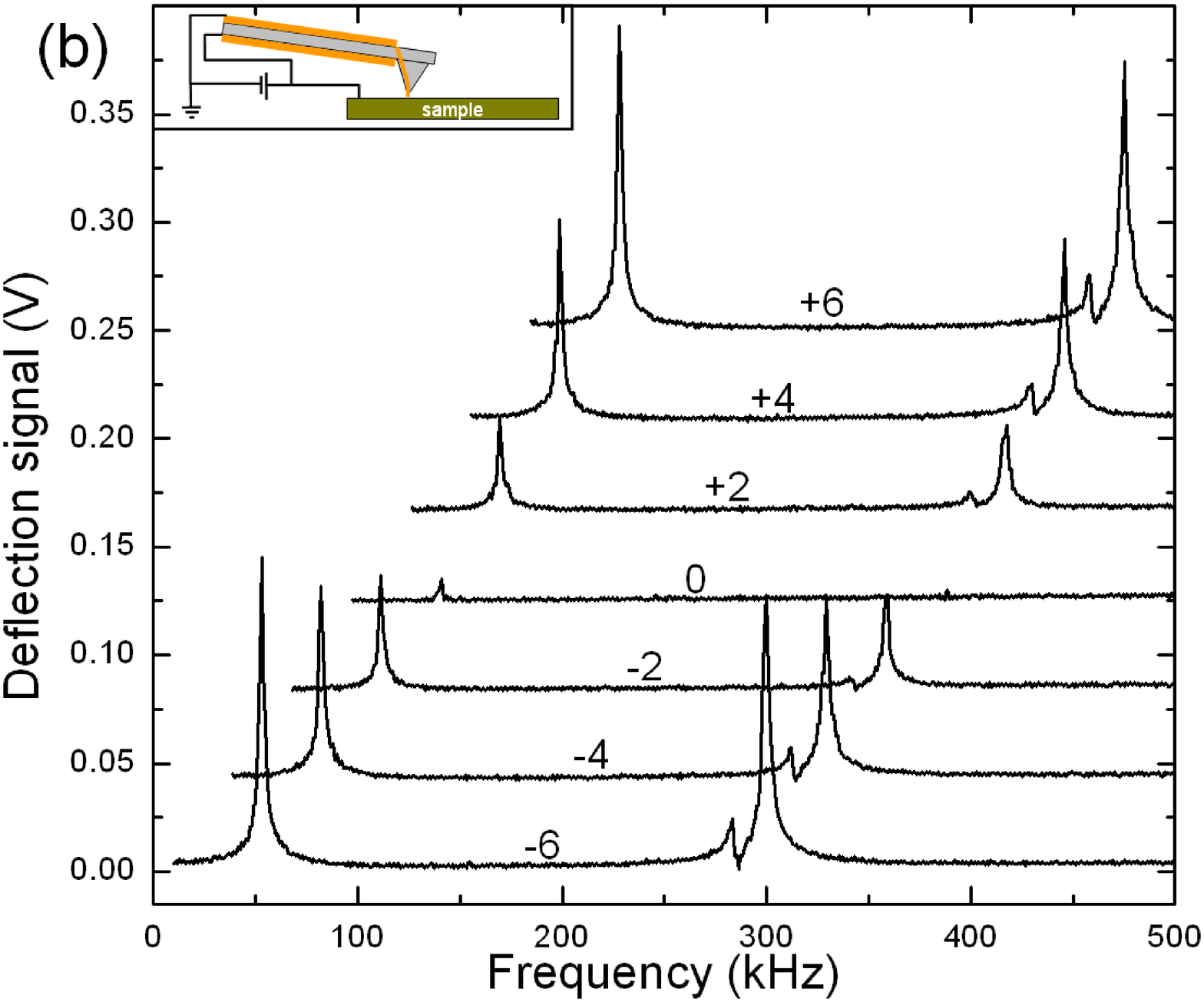}
\end{center}
\caption{Resonance of the probe excited by the electrostatic interaction in 'grounded'(a) and 'ON' (b) configuration at different dc voltages. Curves are offset horizontally and vertically for clarity. Inset of Fig.~2a: comparison between the resonance peak amplitudes vs dc voltage in the two configurations.}
\label{fig2}
\end{figure}

In this Letter we shall demonstrate that it is possible to eliminate the cantilever-to-sample capacitive coupling by using a `ground plane' that we shall call {\it`active shield'} in the following. In Fig.~1a a sketch of this probe is presented together with scanning electron microscope (SEM) pictures of the actual cantilever apex on both sides (Fig.~1b). Our active shield consists of a gold electrode evaporated on the tip-side of the cantilever. FIB nanomachining was employed to eliminate electrical shorts between the top and bottom gold electrodes. The characteristic final resistance between these two electrodes was of the order of a few G$\Omega$. 
All the measurements were performed employing a home-built AFM head and controller, exploiting its absolute positioning stage and force-distance measuring capabilities. Two different triangular cantilevers with different elastic constants were characterized in this work.

In order to study the properties of our probe, we take advantage of the fact that cantilever resonances can be excited by electrostatic interactions\cite{Hong}. In fact, the total attractive force $F_{tot}$ between a cantilever and a conductive sample follows\cite{Abraham2,Baro}:
\[
F_{tot}\propto{\frac{dC}{dz}[V_{dc}^{2}+2V_{dc}V_{ac}\textrm{sin}\omega t+\frac{1}{2}V_{ac}^2(1-\textrm{sin}2\omega t)]}
\]
where $\it{dC/dz}$ is the capacitance gradient and $V_{dc}$ and $V_{ac}$ are dc and ac voltage amplitudes between probe and the sample, respectively. The component of the force at frequency 2$\omega$ depends only on the ac voltage,
while the $\omega$ component depends also on $V_{dc}$. 

In order to excite cantilever mechanical resonances, we connected both electrodes to an ac bias at variable frequency $\omega$ while keeping the sample at fixed dc voltage. This configuration simulates a standard doped cantilever with the removal of the cone-to-sample capacitive interaction (switch position 1 in Fig.~1a). 
The amplitude of the deflection signal at frequency $\omega$ is proportional to $V_{dc}V_{ac}dC/dz$,  it was acquired with a spectrum analyzer as a function of $\omega$ while the probe was lying a few microns above the sample surface. 

The excitation spectrum is reported in Fig.~2. It can be noted that three main resonances are present.
The one at $\sim$54 kHz is the fundamental flexural resonance, those at $\sim$300 kHz are probably related to torsional modes. The resonance-peak amplitude increases linearly with $V_{dc}$, as expected (see also the inset of Fig.~2a). When the active shield is connected to the sample and the ac voltage is applied to the tip only, a dramatic reduction of the resonance is observed. In the following we shall refer to this as the `ON' configuration.
The electrostatic interaction of the biased sample with the tip leads to the expected linear growth of the peak amplitude with the applied dc voltage. A quantitative comparison between the two electrical configurations is reported in the inset of Fig.~2a, where the resonance peak amplitude in both the configuration is shown at different applied dc voltages. We note that in the 'ON' mode the peak amplitude at the resonance is reduced by $\sim$25 dB with respect to the 'grounded' mode. This demonstrates that the active shield effectively reduce the cantilever-to-sample electrostatic interaction allowing at the same time to employ these tips for EFM or KPM measurements.

\begin{figure}[ht!]
\begin{center}
\includegraphics[width=8 cm,clip]{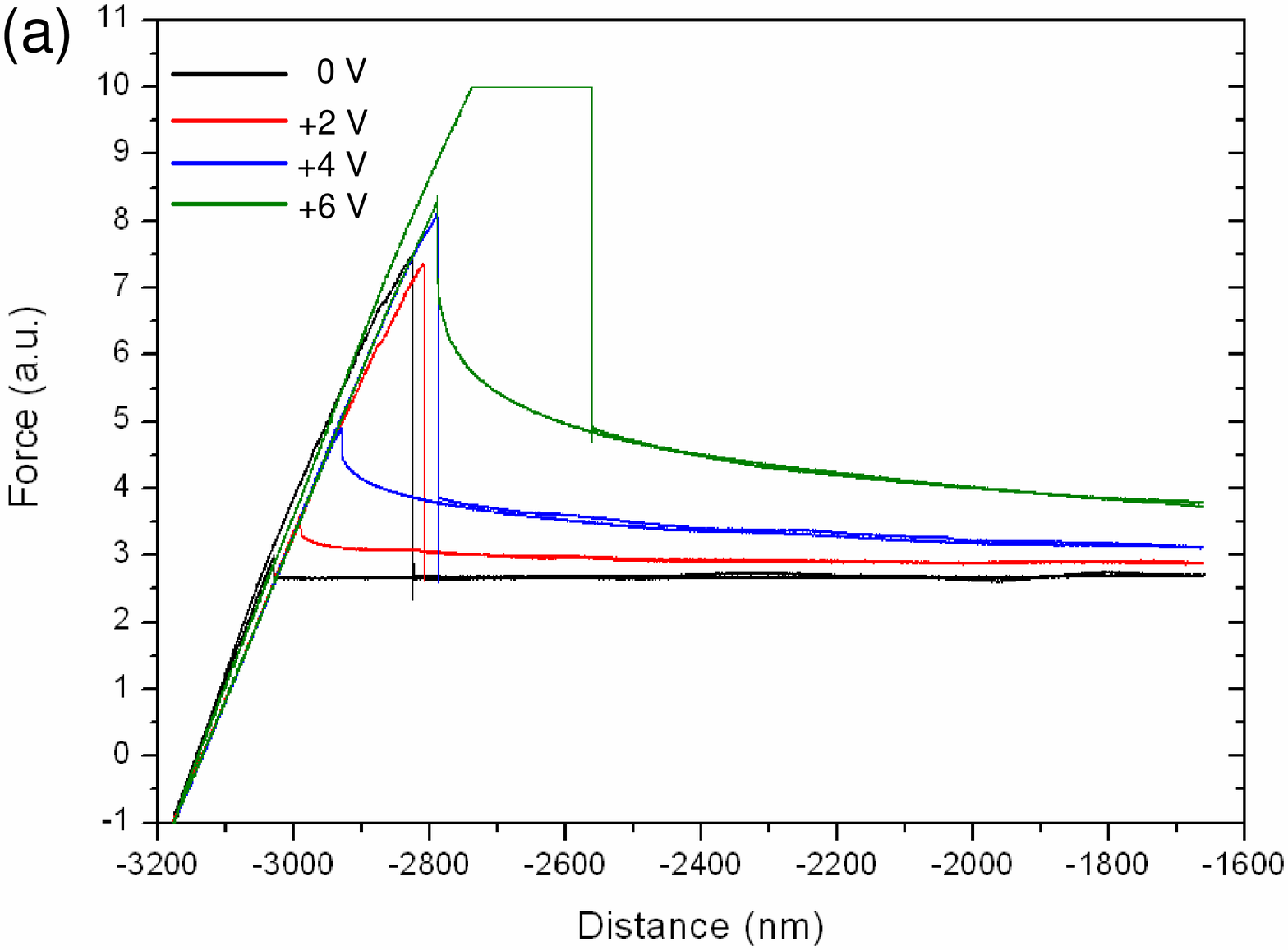}
\includegraphics[width=8 cm,clip]{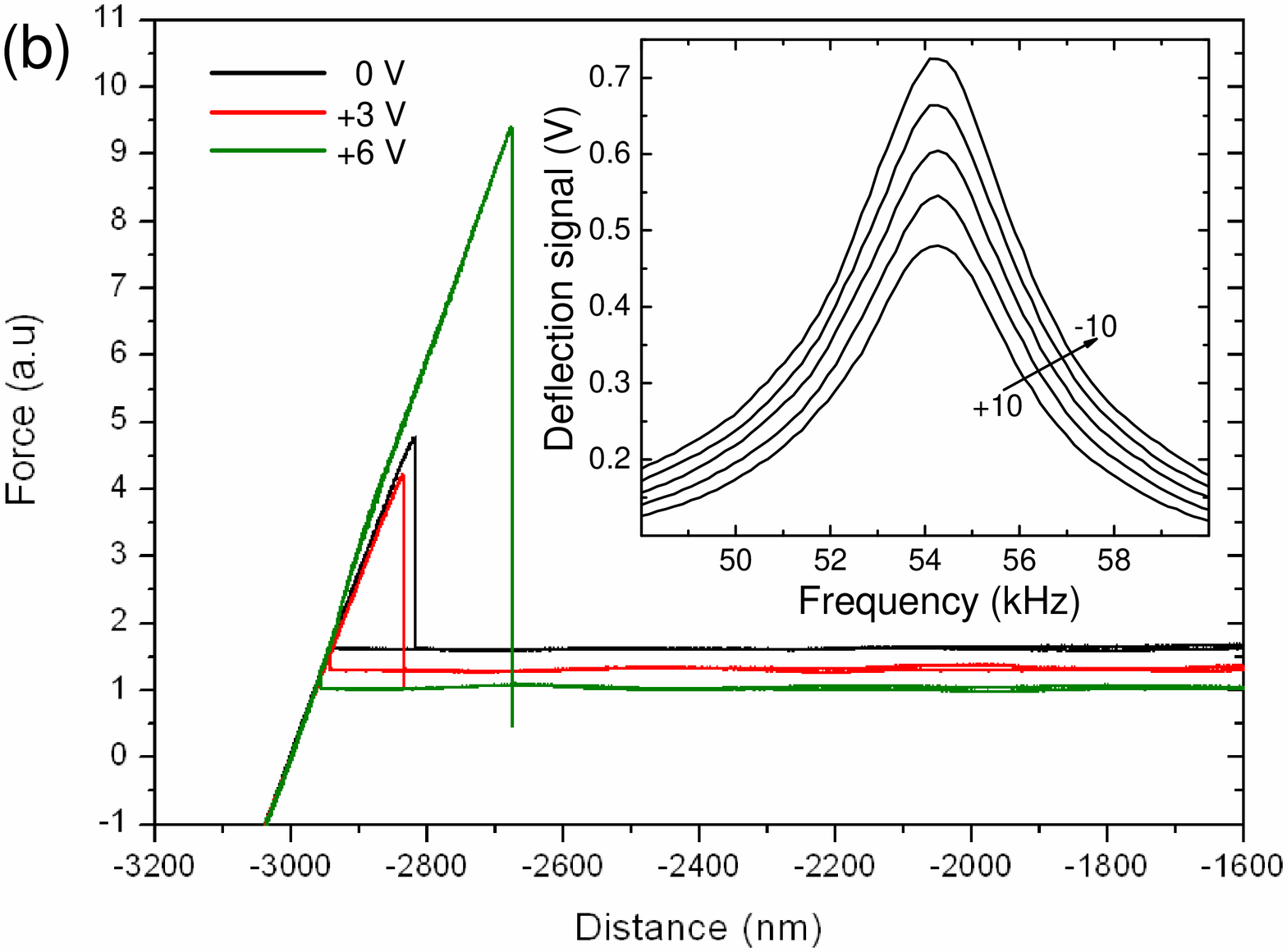}
\includegraphics[width=8 cm,clip]{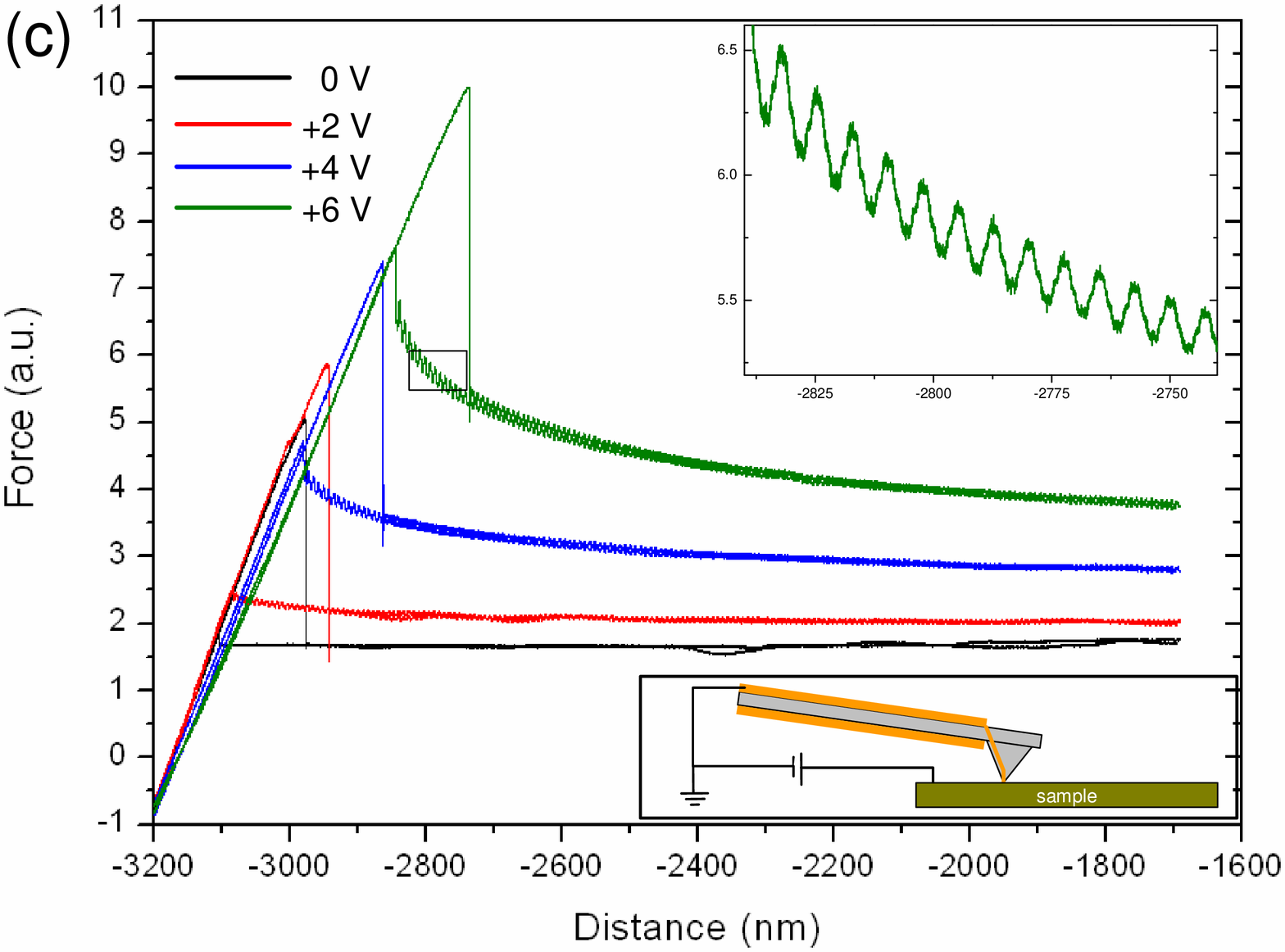}
\end{center}
\caption{(a)Approach and withdrawal FTD curves of the `grounded' probe on a metallic sample (switch position 1);(b) Approach and withdrawal FTD curves of the probe with the `active shield' biased as the sample (switch position 3, ON): in the inset the resonance peak at different dc bias between the two electrodes;(c) Approach and withdrawal FTD curves of the probe having the active shield in 'floating' position: notice the oscillation induced by the line frequency noise (see inset for a detail).}
\label{fig3}
\end{figure}

A force-to-distance (FTD) analysis was also employed to better characterize this probe and to compare its results with a standard cantilever.
Figure 3a reports FTD curves (approach and withdrawal) in a configuration where both electrodes are grounded (switch position 1 in Fig.~1a) and the metallic sample surface is biased at various voltages. In the range of tip-to-sample distance shown the cantilever contribution is very important. This can be inferred by the characteristic behavior of the deflection as a function of distance and applied bias. On the contrary, when the `active shield' is biased at the same potential as the sample (switch position 3 in Fig.~1a) the FTD curves are much less affected by the applied bias (Fig.~3b) with the exception of the jump-to-contact and off-contact regions. In these regions the tip-apex capacitive interaction remains strong and increases with the applied dc voltage. Nevertheless when the voltage is applied the deflection signal at larger distances from the sample indicates a residual voltage dependence corresponding to bending in the opposite direction respect to the other two configurations (see Fig. 3(a) and (c)). This particular behavior does not depend on the distance and is almost linear with the applied bias. 
Studying the peak amplitude of the cantilever resonance as a function of the applied bias between the electrodes we noted that it diminishes linearly (see inset in Fig.~3b) by varying the tip-to-shield bias from negative to positive values. This is consistent with electrostrictive effects on the insulating $Si_{3}N_{4}$ cantilever caused by the biased gold electrodes that cause a mechanical bending of the cantilever and a withdrawal of the tip from the sample surface. Further studies will be carried out to get a deeper understanding of this phenomenon.
When the active shield was left floating we observed a 50 Hz oscillation of the cantilever (see inset in Fig.~3c). Its amplitude was seen to increase with the applied dc bias.
This phenomenon is related to line-frequency noise that couples to the floating active shield giving rise to the $\omega$ component of the electrostatic force. This ac voltage is in any case too low to generate the 2$\omega$ component in the oscillations.

The same general behavior was found when scanning doped GaAs substrates with these probes.

In conclusion, a new probe for applications in electrostatic force microscopy was fabricated by FIB nanolithography starting from a commercially available $Si_{3}N_{4}$ insulating cantilever. We demonstrated that an 'active shield' electrode allowed us to strongly reduce the electrostatic contribution of the cantilever. Its characteristics were analyzed by spectral analysis of electrostatic excitation of its mechanical resonances and by force-distance measurement at different applied dc voltages. The applications of this probe in EFM and KPM were discussed. We believe that this probe can be of interest also in current sensing and LAO where the cantilever contribution to the applied field is important in many circumstances. In fact it reduces the contrast and resolution in the imaging in the first case and contributes with a capacitive current during the lithographic process when an ac bias is applied to the probe in the second case. Moreover, removing the cantilever bending contribution in presence of high humidity is essential in data analysis of water-meniscus formation studies\cite{Perez}. 
Finally we should like to remark that this probe configuration can be easily adapted to commercially available conducting cantilevers.

We should like to thank F. Beltram for useful discussions and careful reading of the manuscript. CM, AA, and PF acknowledge partial financial support for this research by MIUR-FIRB "NOMADE".


\begin{thebibliography}{11}

\bibitem{Terris}
B.D. Terris, J.E. Stern, D. Rugar, and H.J. Mamim {\it Phys. Rev. Lett.} {\bf 63} 2669 (1989);
\bibitem{Abraham1}
D.W. Abraham, C. Williams, J.Slinkman, and H.K. Wichramasinghe, {\it J. Vac. Sci. Tech. B} {\bf 9} 703 (1991);
\bibitem{Abraham2}
J.M.R. Weaver and D.W. Abraham, {\it J. Vac. Sci. Tech. B} {\bf 9} 1559 (1991);
\bibitem{Gomes-Navarro}
C. Gomes-Navarro, A. Gil, M. Alvarez, P.J. de Pablo, F. Moreno-Herrero, I. Horcas, R.Fernandez-Sanchez, J. Colchero, J. Gomez-Herrero, and M. Baro', {\it Nanotechnology} {\bf 13} 314 (2002);
\bibitem{Baro}
J. Colchero, A. Gil, and A.M. Baro', {\it Phys. Rev. B} {\bf 64} 245403 (2001);
\bibitem{Saenz}
G.M. Sacha and J.J. Saenz, {\it Appl. Phys. Lett.} {\bf 85}, 2610 (2004);
\bibitem{Law}
B.M. Law and F. Rieutord, {\it Phys. Rev. B} {\bf 66} 035402 (2002);
\bibitem{Koley}
G. Koley, M.G. Spencer, and H.R. Bhangale, {\it Appl. Phys. Lett.} {\bf 79}, 545 (2001);
\bibitem{Menozzi}
C. Menozzi,G.C. Gazzadi, A. Alessandrini, P. Facci, {\it Ultramicroscopy}, {\bf 104}, 220 (2005);
\bibitem{Hong}
J.W. Hong, Z.G. Khim, A.S. Hou and Sang-il Park, {\it Appl. Phys. Lett.} {\bf 69}, 2831 (1996);
\bibitem{Perez}
J.A. Dagata, F.Perez-Murano, C. Martin, H. Kuramochi and H. Yokoyama, {\it J. of Appl. Phys.} {\bf 96}, 2393 (2004).
\end{thebibliography}
\end{document}